\documentclass[aps,showpacs,nofootinbib,superscriptaddress,twocolumn,showkeys,floatfix]{revtex4-1}

\usepackage[italian,english]{babel}
\usepackage[T1]{fontenc}
\usepackage[utf8x]{inputenc}
\usepackage{xcolor}
\usepackage{soul}

\usepackage{graphicx}
\usepackage{epstopdf}         
\graphicspath{{./Images/}}    

\usepackage{booktabs}

\usepackage{amsmath}   
\usepackage{amssymb}   
\usepackage{bm}        
\usepackage{siunitx}   
\usepackage[version=4]{mhchem}    

\usepackage{datetime}

\makeatletter
\preto\maketitle{%
  \begingroup\lccode`~=`,
  \lowercase{\endgroup
  \let\saved@breqn@active@comma~
  \let~}\active@comma 
}
\appto\maketitle{%
  \begingroup\lccode`~=`,
  \lowercase{\endgroup
  \let~}\saved@breqn@active@comma 
}
\makeatother

\begin{document}
\allowdisplaybreaks[2]

\title{A method for minimizing the magnetic cross-talk\\ in twin-aperture $\cos\theta$ superconducting dipoles}

\author{Alessandro Maria Ricci}
\email{alessandro.ricci@ge.infn.it,\\ alessandromaria.ricci@edu.unige.it}
\affiliation{Dipartimento di Fisica, Università di Genova, via Dodecaneso 33, I-16146 Genova, Italy}
\affiliation{INFN sezione di Genova, via Dodecaneso 33, I-16146 Genova, Italy}

\author{Pasquale Fabbricatore}
\email{pasquale.fabbricatore@ge.infn.it}
\affiliation{INFN sezione di Genova, via Dodecaneso 33, I-16146 Genova, Italy}

\begin{abstract}
We present an analytic method to minimize the magnetic cross-talk in twin-aperture $\cos\theta$ dipoles. In the single-aperture $\cos\theta$ layout, the coil design can be performed with an analytic approach, based on a sector coil approximation. This method allows a fast evaluation of the field harmonics and an almost exhaustive scan on the positions and dimensions of the sectors, for coil layouts made of a different number of sectors. This increases the probabilities to find the coil shape which best fits the specifications. In a twin-aperture arrangement, the magnetic cross-talk can be not negligible and, to the aim of an analytic minimization of the unwanted multipoles, an extension of the single-aperture sector model is required. This is the case of the recombination dipole D2 for the High Luminosity LHC and of the $16$-\si{\tesla} bending dipole for the Future Circular Collider (FCC). This analytical method has been used to find alternative coil designs for both dipoles.
\end{abstract}

\date{\today, \currenttime}

\keywords{Superconducting accelerator magnets, sector coils, cross-talk, field harmonics, High Luminosity Large Hadron Collider (HL-LHC), Future Circular Collider (FCC).}

\maketitle

\section{Introduction}


The superconducting dipoles, bending the particle beams in the high energy accelerators, must provide a high-homogeneous magnetic field. The general rule is that any higher order multipole must be lower than $10^{-4}$ of the central field. Moreover, it is necessary to take into account many constraints on the coil shape (minimum bending radius, maximum magnet dimensions, inter-layer spacers, etc.), the costs, the operating margins, the effects of the persistent currents and the magnetic components, introducing difficulties in the design.

The colliders built so far (Tevatron, HERA, RHIC and LHC) used \ce{NbTi} $\cos\theta$ dipole magnets. In this layout, the coil shape is an annulus and the conductors are piled up in blocks, separated by spacers and carrying the same constant current density~\cite{Mess:Superconducting Accelerator Magnets, Brechna:Superconducting Magnet Systems, Russenschuck:Field Computation}. This practical arrangement aims to approximate an ideal annulus, crossed by a current density proportional to the cosine of the azimuth ($\cos\theta$ annulus). This configuration is the most efficient in the use of superconducting material. However, different coil designs ($\cos\theta$~\cite{McInturff:D20}, common-coil~\cite{Benjegerdes:RD3}, block-coil~\cite{Ferracin:HD2, Milanese:FRESCA2} and canted-$\cos\theta$~\cite{Montenegro:CCT}) were tested in \ce{Nb3Sn} to overcome manufacturing challenges and manage the stress. A summary for \ce{Nb3Sn} superconducting dipoles can be found in Ref.~\cite{Schoerling:Nb3Sn magnets}.

In the colliders two particle beams are counter-rotating and require two separate channels with opposite magnetic fields. Tevatron, HERA and RHIC have two separate storage rings and the superconducting magnets are designed with the beam pipe surrounded by the coil inside the iron yoke (single-aperture layout). The available space in the LHC tunnel does not allow two separate storage rings. Thus, the main dipoles (MB), the main quadrupoles (MQ) and some magnets for the corrections, the insertion regions and the interaction regions are designed with the two beam pipes surrounded by the coils inside a common iron yoke~\cite{CERN:LHC, Russenschuck:Field Computation} (twin-aperture layout). In this configuration, each coil can generate unwanted multipoles in the other aperture (cross-talk). This problem is particularly important for a special class of dipoles involved in proximity of the collider Interaction Regions (IR), the recombination dipoles D2~\cite{CERN:LHC, Russenschuck:Field Computation}. These special magnets are used for recombining the beams before the collision in the Interaction Point (IP). In order to achieve this, the magnetic field must have the same polarity in both apertures. In the LHC dipole D2, the $2.77$-\si{\tesla} magnetic field at the $188$-\si{\milli\meter} inter-beam distance is high enough for generating a non-negligible cross-talk, but low enough for allowing the iron yoke to magnetically decouple the coils. However, this solution is no more viable in the new dipole D2 for the High Luminosity upgrade of LHC (HL-LHC)~\cite{Farinon:D2, Fabbricatore:D2, Bersani:D2, CERN:HL-LHC}, because the higher magnetic field ($4.5$~\si{\tesla}) saturates the iron yoke, resulting in a dramatic increase of the unwanted multipoles. Therefore, the iron yoke between the coils has been removed and the field quality is tuned by an asymmetric coil winding.

Recently, the Future Circular Collider Study (FCC) published the Conceptual Design Report (CDR)~\cite{FCC Study:CDR Vol. 1,FCC Study:CDR Vol. 2, FCC Study:CDR Vol. 3, FCC Study:CDR Vol. 4}, which describes the feasibility of high-performance colliders, housed in a new 100-km tunnel in the area of Geneva. The hadron collider (FCC-hh) would achieve a $100$-\si{\tera\electronvolt} collision energy and its \ce{Nb3Sn} bending dipoles would generate a $16$-\si{\tesla} magnetic field. Alternatively, the $16$-\si{\tesla} magnets could be used for a High Energy upgrade of the Large Hadron Collider (HE-LHC), which would increase the collision energy from $14$~\si{\tera\electronvolt} to $27$~\si{\tera\electronvolt}. The EuroCirCol Collaboration studied different designs~\cite{Tommasini:EuroCirCol 1, Tommasini:EuroCirCol 2, Schoerling:EuroCirCol, Lorin:16T block-coil 1, Lorin:16T block-coil 2, Segreti:16T block-coil, Toral:16T common-coil 1, Toral:16T common-coil 2, Auchmann:16T canted-cos-theta, Montenero:16T canted-cos-theta, Sorbi:16T cos-theta, Marinozzi:16T cos-theta, Caiffi:16T cos-theta, Valente:16T cos-theta} for the $16$-\si{\tesla} superconducting dipoles and chose as baseline the $\cos\theta$ layout~\cite{Schoerling:EuroCirCol, Valente:16T cos-theta}, whose coil cross-section is left-right asymmetric, because in the cold mass size constraint the elevate magnetic field leads to a strong cross-talk, which can be controlled only by an asymmetric coil winding.

In this paper we present an analytic method, which minimizes the magnetic cross-talk by finding the asymmetric coil cross-sections. The first step in the coil design is to find the block arrangements, which generate a high-homogeneous magnetic field given the bending radius, the cable width, the layer number and the inter-beam distance. These configurations cannot be derived explicitly and many numerical algorithms exist to find the optimal cross-sections~\cite{Russenschuck:Field Computation}. Owing to the complicated magnet geometry (coils made of blocks, blocks made of cables and cables made of strands), they are time-consuming and, to be really effective, they have to operate on configurations which are not too far from a local optimum. Therefore, analytical models approximating the blocks as annular sectors (sector coil models)~\cite{Mess:Superconducting Accelerator Magnets, Brechna:Superconducting Magnet Systems, Bailey:sector coil, Devred:sc magnets} can be used to carry out an initial scan on a very large number of possible configurations~\cite{Bailey:sector coil, Borgnolutti:HL-LHC quadrupoles, Louzguiti:FCC sextupoles and octupoles}. This makes easier to find the cross-section which best suits the specifications. We extended the current sector model to analytically describe the contribution to the harmonic components, which one coil exerts on the other aperture. We used this extended sector model to find the asymmetric coil configurations, which minimize the cross-talk in the new dipole D2 of HL-LHC and in the $16$-\si{\tesla} bending dipole of FCC. In a computational time of few minutes for D2 and of few tens of minutes for the FCC dipole, this method allowed to find alternative magnetic designs, which have an excellent field quality. These results show that this method can be used as a complementary tool at the early stage of the coil design of a twin-aperture dipole, which presents a non-negligible cross-talk.

In Section II we review the current sector model, in Section III we explain the extended sector model and the resolving procedure, finally in Section IV we show the results.

\section{The current sector model}

In the complex formalism, Biot and Savart's law set that a current line $I$ in the position $z_0\equiv x_0+iy_0$ generates a magnetic field $B(z)\equiv B_y(z)+iB_x(z)$ in the position $z\equiv x+iy$ according to the formula
\begin{equation}
    B\left(z\right)=\frac{\mu_0 I}{2\pi\left(z-z_0\right)} \,.
\end{equation}
Knowing that for $|z|<1$
\begin{equation}
    \frac{1}{1-z}=1+z+z^2+z^3+\dots=\sum_{n=1}^{\infty}z^{n-1} \,,
\end{equation}
we can develop the multipolar expansion of the magnetic field for $|z|<|z_0|$, as
\begin{equation}
\begin{split}
    B\left(z\right) & =\frac{\mu_0 I}{2\pi\left(z-z_0\right)}=-\frac{\mu_0 I}{2\pi z_0}\frac{1}{1-z/z_0}  \\[0.3cm]
    & =-\frac{\mu_0 I}{2\pi z_0}\sum_{n=1}^{\infty}\left(\frac{z}{z_0}\right)^{n-1}  \\[0.3cm]
    & =-\frac{\mu_0 I}{2\pi z_0}\sum_{n=1}^{\infty}\left(\frac{R_{ref}}{z_0}\right)^{n-1}\left(\frac{z}{R_{ref}}\right)^{n-1},
\end{split}
\label{eq:multipolar expansion}
\end{equation}
where $R_{ref}$ is a reference radius usually chosen as $2/3$ of the aperture radius. We can re-write the multipolar expansion in the European notation as
\begin{equation}
\begin{split}
    B\left(x,y\right)=B_y+iB_x=\sum_{n=1}^\infty\left(B_n+iA_n\right)\left(\frac{x+iy}{R_{ref}}\right)^{n-1} \,,
\end{split}
\label{eq:complex magnetic field}
\end{equation}
where
\begin{equation}
\begin{split}
    B_n+iA_n & =-\frac{\mu_0 I}{2\pi z_0}\left(\frac{R_{ref}}{z_0}\right)^{n-1}  \\[0.3cm]
    & =-\frac{\mu_0 I}{2\pi R_{ref}}\left(\frac{R_{ref}}{z_0}\right)^{n} \,.
\end{split}
\label{eq:cylindrical harmonics}
\end{equation}
The coefficients $A_n$ and $B_n$ have the dimensions of the magnetic field (\si{\tesla}) and they are called skew and normal cylindrical harmonics respectively. In the European definition~\eqref{eq:complex magnetic field}, each component of order $n$ represents the $2n$-pole component. The cylindrical harmonics of a dipole can be dimensionless and normalized to units as $b_n=10^{4}B_n/B_1$ and $a_n=10^{4}A_n/B_1$, where $B_1$ is the dipole component generated from a current line, which follows from Eq.~\eqref{eq:cylindrical harmonics}
\begin{equation}
    B_1=-\frac{\mu_0 I}{2\pi}Re\left(\frac{1}{z_0}\right)=-\frac{\mu_0 I}{2\pi}\frac{x_0}{x_0^2+y_0^2} \,.
\end{equation}
Thus Eq.~\eqref{eq:complex magnetic field} becomes
\begin{equation}
    B_y+iB_x=10^{-4}B_1\sum_{n=1}^\infty\left(b_n+ia_n\right)\left(\frac{x+iy}{R_{ref}}\right)^{n-1} \,,
\end{equation}
where $a_n$ and $b_n$ are called normalized skew and normal cylindrical harmonics respectively.

\begin{figure}[t]
\includegraphics[width=0.38\textwidth]{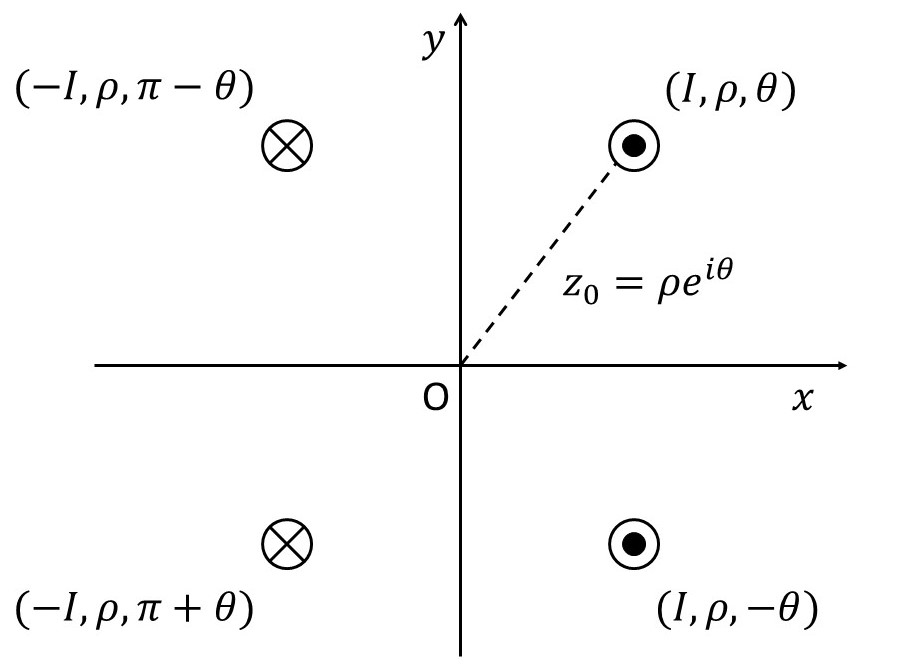}
\caption{Quadruplet of current lines with an even symmetry about the $x$-axis and an odd symmetry about the $y$-axis.}
\label{fig:symmetric current-line quadruplet}
\end{figure}

Now let us consider the quadruplet of current lines $(I,\rho,\theta),(-I,\rho,\pi-\theta),(-I,\rho,\pi+\theta),(I,\rho,-\theta)$ shown in Fig.~\ref{fig:symmetric current-line quadruplet}. The magnetic field generated by this quadruplet is the sum of the contributions of each current line and, within the circle of center $O$ and radius $\rho$, it can be calculated by Eq.~\eqref{eq:complex magnetic field}, where
\begin{equation}
\begin{split}
    B_n+ & iA_n=-\frac{\mu_0 I}{2\pi R_{ref}}\left(\frac{R_{ref}}{\rho}\right)^n \\[0.3cm]
    & \times \left(e^{-in\theta}-e^{-in\left(\pi-\theta\right)}-e^{-in\left(\pi+\theta\right)}+e^{in\theta}\right) \,.
\end{split}
\end{equation}
Knowing that
\begin{equation}
\begin{split}
    e^{-in\theta}-e^{-in\left(\pi-\theta\right)} & -e^{-in\left(\pi+\theta\right)}+e^{in\theta} \\[0.3cm]
    & =2\left[1-\left(-1\right)^n\right]\cos n\theta \,,
\end{split}
\end{equation}
which is only non-zero when $n$ is odd, the complex magnetic field can be written for $|z|<\rho$ with the only non-zero coefficients
\begin{equation}
    B_{n}=\frac{2\mu_0 I}{\pi R_{ref}}\left(\frac{R_{ref}}{\rho}\right)^{n}\cos n\theta \quad n\;\text{odd}\,,
\label{eq:allowed harmonics}
\end{equation}
which are called allowed cylindrical harmonics of this current distribution. The even symmetry about the $x$-axis deletes the skew harmonics and the odd symmetry about the $y$-axis drops the even normal harmonics.

\begin{figure}[t]
\includegraphics[width=0.38\textwidth]{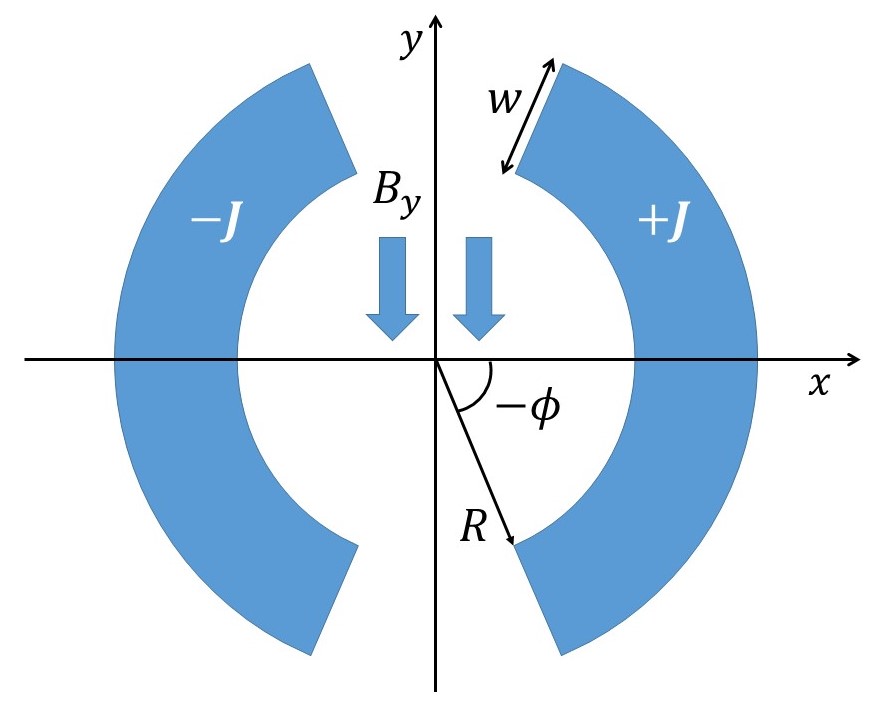}
\caption{Sector coil layout for a dipole of inner radius $R$ and coil width $w$, spanning the angle from $0$ to $\phi$. The layout is symmetric both about the $x$-axis and about the $y$-axis. In the right half coil a uniform current density $J$ flows and in the left $-J$. The magnetic field for $r<R$ has the component $B_y$ only.}
\label{fig:sector dipole}
\end{figure}

Let us consider a dipole, whose quarter coil layout is a sector of width $w$ and bending radius $R$, spanning the angle from $0$ to $\phi$. The layout is symmetric both about the $x$-axis and about the $y$-axis. A uniform current density $J$ flows in the right half coil and $-J$ in the left (see Fig.~\ref{fig:sector dipole}). The allowed harmonics can be obtained by integrating the current in Eq.~\eqref{eq:allowed harmonics} over the sector:
\begin{equation}
B_n=\frac{2\mu_0 J R_{ref}^{n-1}}{\pi n\left(n-2\right)}\left(\frac{1}{\left(R+w\right)^{n-2}}-\frac{1}{R^{n-2}}\right)\sin n\phi \,.
\label{eq:harmonics for one sector}
\end{equation}
Solving Eq.~\eqref{eq:harmonics for one sector} we can find the angles that set to zero the first allowed harmonics $B_3$ and the ones that cancel the second allowed harmonics $B_5$. Because the angles are different, we cannot have $B_3=B_5=0$ with a single sector. If we consider a coil composed by two sectors $[0,\phi_1]$ and $[\phi_2,\phi_3]$, with a wedge between $\phi_1$ and $\phi_2$, we can set $B_3=B_5=B_7=0$ by numerically solving the equation system
\begin{align}
    & \sin 7\phi_3 - \sin 7\phi_2 + \sin 7\phi_1 =0 \,, \\[0.3cm]
    & \sin 5\phi_3 - \sin 5\phi_2 + \sin 5\phi_1 =0 \,, \\[0.3cm]
    & \sin 3\phi_3 - \sin 3\phi_2 + \sin 3\phi_1 =0 \,.
\end{align}

At a early stage of the design process, we can use the sector coils to search the minimum number, the positions and the dimensions of the sector blocks, which meet the field quality requirements. This analytical model allows a very fast scan on a very large number of possible configurations.

\section{The extended sector model}

\begin{figure}[t]
\includegraphics[width=0.38\textwidth]{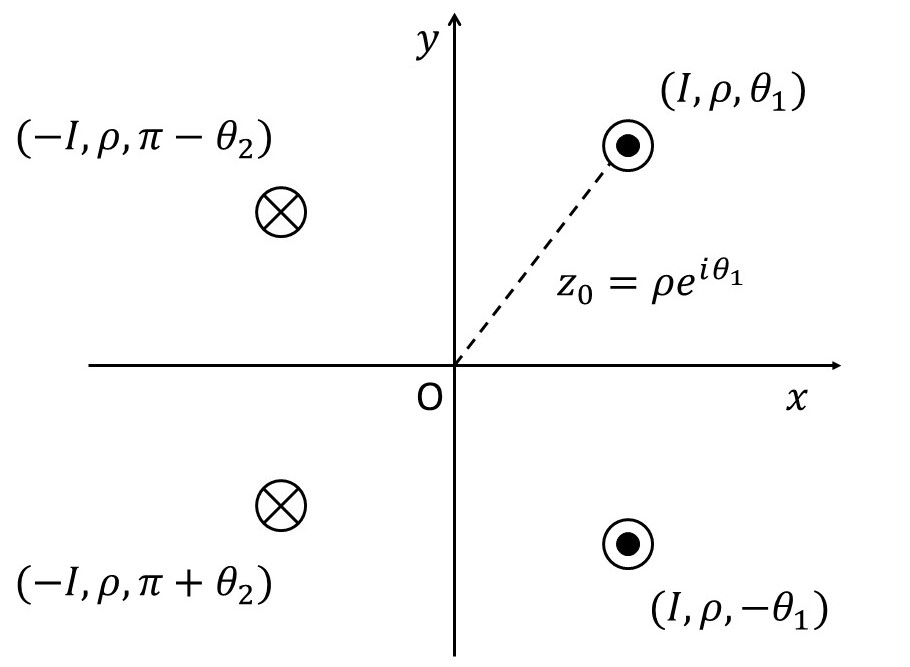}
\caption{Asymmetric quadruplet of current lines about the $y$-axis and with an even symmetry about the $x$-axis.}
\label{fig:asymmetric current-line quadruplet}
\end{figure}

The twin-aperture layout introduces a complicating factor, i.e. the contribution to the harmonic components which one coil exerts on the other aperture. This cross-talk can bring to non-zero normal coefficients also for even orders. To control the even normal harmonics we must break the symmetry of the current lines about the $y$-axis. Therefore, we consider a quadruplet of current lines, which is symmetric only about the $x$-axis (see Fig.~\ref{fig:asymmetric current-line quadruplet}). The normal multipoles can be written as
\begin{equation}
\begin{split}
    B_n= & -\frac{\mu_0 I}{\pi R_{ref}}\left(\frac{R_{ref}}{\rho}\right)^n \cos n\theta_1  \\[0.3cm]
    & -\frac{\mu_0 \left(-I\right)}{2\pi R_{ref}}\left(\frac{R_{ref}}{\rho}\right)^n \cos n\left(\pi-\theta_2\right) \\[0.3cm]
    & -\frac{\mu_0 \left(-I\right)}{2\pi R_{ref}}\left(\frac{R_{ref}}{\rho}\right)^n \cos n\left(\pi+\theta_2\right) \,.
\end{split}
\label{eq:asymmetric current-line quadruplet}
\end{equation}
Knowing that
\begin{equation}
\begin{aligned}
    & \cos n\left(\pi-\theta_2\right)=\left(-1\right)^n \cos n\theta_2 \,, \\[0.3cm]
    & \cos n\left(\pi+\theta_2\right)=\left(-1\right)^n \cos n\theta_2 \,,
\end{aligned}
\label{eq:cosine}
\end{equation}
Eq.~\eqref{eq:asymmetric current-line quadruplet} becomes
\begin{equation}
\begin{split}
    B_n= & -\frac{\mu_0 I}{\pi R_{ref}}\left(\frac{R_{ref}}{\rho}\right)^n \\[0.3cm]
    & \quad \times \bigl[\cos n\theta_1 - \left(-1\right)^n \cos n\theta_2\bigr] \,.
\end{split}
\end{equation}
Integrating the current for passing to an asymmetric sector coil about the $y$-axis, we obtain
\begin{equation}
\begin{split}
    B_n & =-\frac{\mu_0 J R_{ref}^{n-1}}{\pi} \int_R^{R+w}\frac{1}{\rho^{n-1}}d\rho  \\[0.3cm]
    & \times\left(\int_{\phi}^{\phi^{'}}\cos n\theta_1 d\theta_1-\left(-1\right)^n\int_{\psi}^{\psi^{'}}\cos n\theta_2d\theta_2\right) \,,
\end{split}
\label{eq:integration of asymmetric current lines}
\end{equation}
where $\phi$ and $\phi^{'}$ are the starting and final angles respectively for the right sector and $\psi$ and $\psi^{'}$ are the starting and final angles respectively for the left sector. We get for $n\neq 2$
\begin{equation}
\begin{split}
    B_n & =\frac{\mu_0 J R_{ref}^{n-1}}{\pi n\left(n-2\right)}\left(\frac{1}{\left(R+w\right)^{n-2}}-\frac{1}{R^{n-2}}\right) \\[0.3cm]
    & \times\left[\sin n\phi^{'}-\sin n\phi-\left(-1\right)^n \left(\sin n\psi^{'}-\sin n\psi\right)\right] \,,
\end{split}
\label{eq:asymmetric sector coil n}
\end{equation}
and for $n=2$
\begin{equation}
\begin{split}
    B_2 & =\frac{\mu_0 J R_{ref}}{2\pi}\ln{\frac{R}{R+w}}  \\[0.3cm]
    & \times\left(\sin 2\phi^{'}-\sin 2\phi-\sin 2\psi^{'}+\sin 2\psi\right) \,.
\end{split}
\label{eq:asymmetric sector coil 2}
\end{equation}
\begin{figure}[t]
\includegraphics[width=0.38\textwidth]{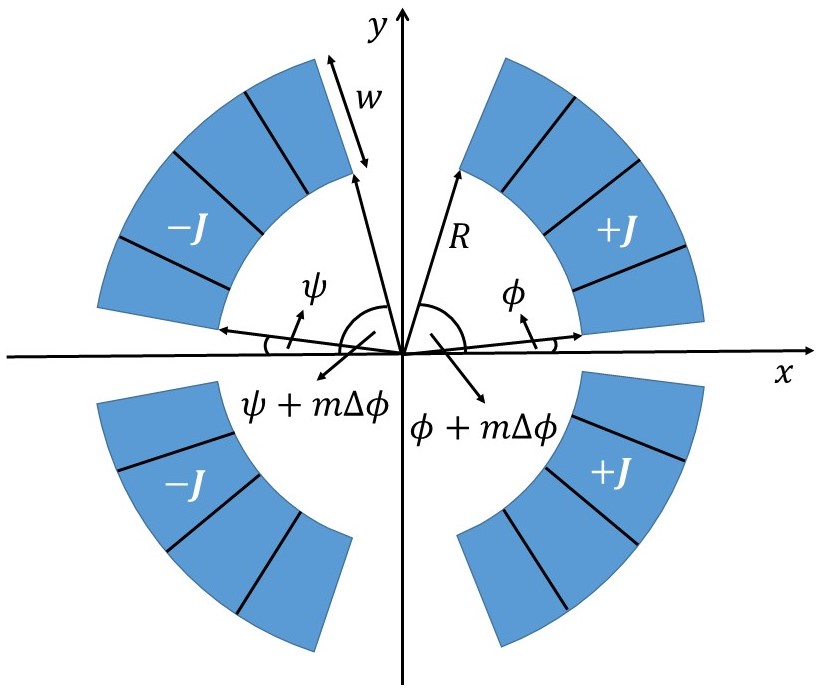}
\caption{Asymmetric sector coil about the $y$-axis and symmetric about the $x$-axis. $\phi$ and $\psi$ are the starting angles of the right and left sectors respectively. Both sectors have the bending radius $R$ and the width $w$. They are composed from $m$ turns and $\Delta\phi$ is the angle underlying each turn.}
\label{fig:asymmetric sector coil}
\end{figure}

For including the discrete size of the cable in the sector model, we define the final angles as
\begin{equation}
\begin{aligned}
    & \phi^{'}=\phi + m\, \Delta\phi \,, \\[0.3cm]
    & \psi^{'}=\psi + m\, \Delta\phi \,,
\end{aligned}
\end{equation}
where $m$ is the number of turns of each sector and $\Delta\phi$ is the angle underlying the turn (see Fig.~\ref{fig:asymmetric sector coil}). It is calculated as
\begin{equation}
\Delta\phi=\arcsin\frac{\bar{l}}{\bar{R}} \,,
\end{equation}
where $\bar{l}$ is the mean cable thickness, considered as conductor plus insulation, and $\bar{R}=R+w/2$ is the mean bending radius of the cable. Therefore, the equations~\eqref{eq:asymmetric sector coil n} and~\eqref{eq:asymmetric sector coil 2} are rewritten as
\begin{equation}
\begin{split}
    B_n & =\frac{\mu_0 J R_{ref}^{n-1}}{\pi n\left(n-2\right)}\left(\frac{1}{\left(R+w\right)^{n-2}}-\frac{1}{R^{n-2}}\right) \\[0.3cm]
    & \quad \times\bigg[\sin n\left(\phi+m\, \Delta\phi\right)-\sin n\phi  \\[0.3cm]
    & \qquad -\left(-1\right)^n \left(\sin n\left(\psi+m\, \Delta\phi\right)-\sin n\psi\right)\bigg]
\end{split}
\label{eq:final asymmetric sector coil n}
\end{equation}
and
\begin{equation}
\begin{split}
    B_2 & =\frac{\mu_0 J R_{ref}}{2\pi}\ln{\frac{R}{R+w}}  \\[0.3cm]
    & \times\bigg[\sin 2\left(\phi + m\, \Delta\phi\right)-\sin 2\phi  \\[0.3cm]
    & \qquad -\sin 2\left(\psi + m\, \Delta\phi\right)+\sin 2\psi\bigg] \,.
\end{split}
\label{eq:final asymmetric sector coil 2}
\end{equation}

The choice to work with the mean cable thickness on the mean bending radius has been done to minimize the harmonic error, due to the geometric differences between the sector coil and the real coil, which are not negligible in the FCC dipole. The maximum error on the most sensible harmonic becomes $\Delta b_3<19$ units in the $16$-\si{\tesla} bending dipole and $\Delta b_3<11$ units in the dipole D2. In the twin-aperture layout, Eq.~\eqref{eq:final asymmetric sector coil n} and Eq.~\eqref{eq:final asymmetric sector coil 2} are the harmonics generated from the right coil in the right aperture and, hereafter, we indicate them as $B_n^r$.

\begin{figure*}[t]
\includegraphics[width=0.7\textwidth]{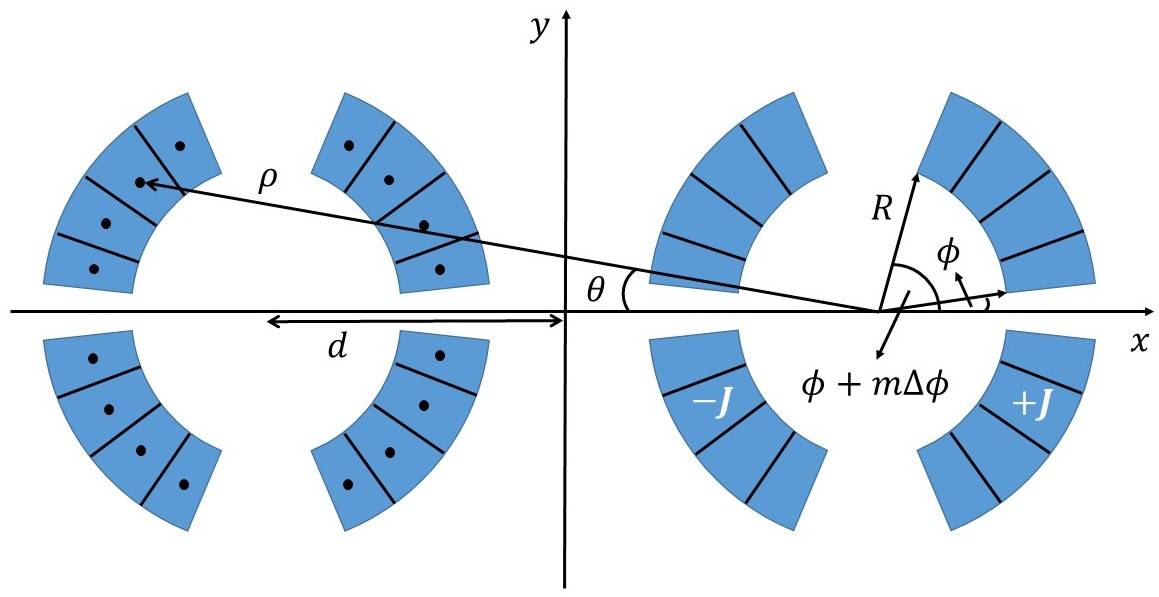}
\caption{Twin-aperture layout. The two sector coils are mirrored, because the beams are counter-rotating. Each coil has the asymmetric cross-section about its vertical axis. The conductors of the left coil are approximated by single current lines, flowing in the center of the conductors themselves (black points); $\rho$ and $\theta$ are the coordinates of each current line. $d$ is half of the inter-beam distance.}
\label{fig:twin-aperture layout}
\end{figure*}

The left coil conductors are far from the right coil aperture, the region where the harmonics are computed. Therefore, we can analytically describe the left coil harmonic contribution approximating each conductor by a single current line, flowing in the center of the turn itself (see Fig.~\ref{fig:twin-aperture layout}). This approximation brings to a maximum error of about $5\%$.

The contribution of the left coil, in the right aperture, is
\begin{equation}
\begin{split}
    B_n^l & =-\frac{\mu_0 I R_{ref}^{n-1}}{2\pi}\sum_{i=1}^m \frac{\cos n \left(\pi-\theta_i\right)}{\rho_i^n}  \\[0.3cm]
    & \qquad -\frac{\mu_0 I R_{ref}^{n-1}}{2\pi}\sum_{i=1}^m \frac{\cos n \left(\pi+\theta_i\right)}{\rho_i^n}  \\[0.3cm]
    & \qquad -\frac{\mu_0 \left(-I\right) R_{ref}^{n-1}}{2\pi}\sum_{i=1}^m \frac{\cos n \left(\pi-\theta_i^{'}\right)}{\left(\rho_i^{'}\right)^n} \\[0.3cm]
    & \qquad \qquad -\frac{\mu_0 \left(-I\right) R_{ref}^{n-1}}{2\pi}\sum_{i=1}^m \frac{\cos n \left(\pi+\theta_i^{'}\right)}{\left(\rho_i^{'}\right)^n}  \,,
\end{split}
\end{equation}
where $\rho_i$ and $\theta_i$ are the polar coordinates of the current lines of the left sectors and $\rho_i^{'}$ and $\theta_i^{'}$ are the polar coordinates of the current lines of the right sectors. By using Eq.~\eqref{eq:cosine}, we get
\begin{equation}
\begin{split}
    B_n^l & =-\left(-1\right)^n\frac{\mu_0 I R_{ref}^{n-1}}{\pi}\sum_{i=1}^m \frac{\cos n\theta_i}{\rho_i^n} \\[0.3cm]
    &\qquad -\left(-1\right)^n\frac{\mu_0 \left(-I\right) R_{ref}^{n-1}}{\pi}\sum_{i=1}^m \frac{\cos n\theta_i^{'}}{\left(\rho_i^{'}\right)^n} \,.
\end{split}
\label{eq:left coil harmonics}
\end{equation}

We set the versus of the currents in way that the magnetic field in the left aperture has polarity opposite than the right aperture. This is the case of the $16$-\si{\tesla} bending dipole for FCC. Instead, for the recombination dipole D2 of HL-LHC, the magnetic field must have the same polarity in the two aperture and this condition is realized if we reverse the versus of the currents in Eq.~\eqref{eq:left coil harmonics}. The current density $J$, which flows in the conductors of the right coil, is linked to the current intensity $I$, as $J=I/S$, where $S$ is the area of each conductor, computed as
\begin{equation}
    S=\frac{\left(R+w\right)^2-R^2}{2}\Delta\phi \,.
\end{equation}

The polar coordinates of the current lines, $\rho_i,\theta_i,\rho_i^{'},\theta_i^{'}$, are linked to the angles $\phi$ or $\psi$ of the corresponding sector of the right coil, by the simple trigonometric formulas. First, we define the polar coordinates of the current lines in the middle of each turn of the right coil as
\begin{equation}
\begin{aligned}
&r=R+\frac{w}{2} \,, \\[0.3cm]
&\gamma_{i}=\phi+\left(i+\frac{1}{2}\right)\Delta\phi \,, \\[0.3cm]
&\gamma_{i}^{'}=\psi+\left(i+\frac{1}{2}\right)\Delta\phi \,,
\end{aligned}
\label{eq:trigonometry}
\end{equation}
where $i$ is an integer number from $0$ to $m-1$. Then, we set the polar coordinates of the current lines of the left coil, splitting between right and left sectors. Indeed, because the beams are counter-rotating, the left coil is mirrored to the right coil. Therefore, the left sectors of the left coil correspond to the right sectors of the right coil and the right sectors of the left coil correspond to the left sectors of the right coil.

For the left sectors of the left coil, the trigonometric formulas are
\begin{equation}
\begin{aligned}
&\theta_{i}=\arctan\left(\frac{r\sin\gamma_{i}}{2d+r\cos\gamma_{i}}\right) \,, \\[0.3cm]
&\rho_{i}=\frac{2d+r\cos\gamma_{i}}{\cos\theta_{i}} \,,
\end{aligned}
\label{eq:external trigonometry}
\end{equation}
where $d$ is half of the inter-beam distance; while for the right sectors of the left coil we have
\begin{equation}
\begin{aligned}
&\theta_{i}^{'}=\arctan\left(\frac{r\sin\gamma_{i}^{'}}{2d-r\cos\gamma_{i}^{'}}\right) \,, \\[0.3cm]
&\rho_{i}^{'}=\frac{2d-r\cos\gamma_{i}^{'}}{\cos\theta_{i}^{'}} \,.
\end{aligned}
\label{eq:internal trigonometry}
\end{equation}

Equations~\eqref{eq:trigonometry},~\eqref{eq:external trigonometry} and~\eqref{eq:internal trigonometry} require to rewrite the summation in Eq.~\eqref{eq:left coil harmonics}, which becomes
\begin{equation}
\begin{split}
    B_n^l & =-\left(-1\right)^n\frac{\mu_0 I R_{ref}^{n-1}}{\pi}\sum_{i=0}^{m-1} \frac{\cos n\theta_i}{\rho_i^n} \\[0.3cm]
    &\qquad -\left(-1\right)^n\frac{\mu_0 \left(-I\right) R_{ref}^{n-1}}{\pi}\sum_{i=0}^{m-1} \frac{\cos n\theta_i^{'}}{\left(\rho_i^{'}\right)^n} \,.
\end{split}
\label{eq:left coil harmonics 2}
\end{equation}

The normalized harmonics, produced in the right coil aperture by the two coils, are
\begin{equation}
\begin{split}
    & b_n^{coil}\left(\phi_1, \psi_1, m_1, \dots, \phi_N, \psi_N, m_N\right) \\[0.3cm]
    & \; =10^4\frac{\sum_{p=1}^N \bigl[B_n^r\left(\phi_p, \psi_p, m_p\right) + B_n^l\left(\phi_p, \psi_p, m_p\right)\bigr]}{\sum_{p=1}^N \bigl[B_1^r\left(\phi_p, \psi_p, m_p\right) + B_1^l\left(\phi_p, \psi_p, m_p\right)\bigr]} \, ,
\end{split}
\label{eq:normalized coil harmonics}
\end{equation}
where $N$ is the number of the coil sectors.

Now, we must consider the iron yoke saturation and the harmonic error due to the geometric differences between the sector coil and the real coil. We observed that these contributions poorly depend from the ``coordinates'' of the configuration $(\phi_p,\psi_p,m_p)$. Then, we can regard the shift from the model solution, $\Delta b_n^{sat+geom}$, approximately as a constant value, which can be estimated by a single FEM evaluation~\cite{Russenschuck:Field Computation}. Therefore, the total normalized harmonics are
\begin{equation}
\begin{split}
    & b_n\left(\phi_1, \psi_1, m_1, \dots, \phi_N, \psi_N, m_N\right) \\[0.3cm]
    & \; =b_n^{coil}\left(\phi_1, \psi_1, m_1, \dots, \phi_N, \psi_N, m_N\right) + \Delta b_n^{sat+geom} \,.
\end{split}
\label{eq:total normalized harmonics}
\end{equation}

The quadratic sum of the total normalized harmonics is minimized by an iterative method:
\begin{enumerate}
    \item we generate a random configuration of the left coil $\left(\phi_1, \psi_1, m_1, \dots, \phi_N, \psi_N, m_N\right)$ and compute the harmonics $B_n^l\left(\phi_p, \psi_p, m_p\right)$ in Eq.~\eqref{eq:normalized coil harmonics};
    \item we numerically find a configuration of the right coil $\left(\phi_1^{'}, \psi_1^{'}, m_1^{'}, \dots, \phi_N^{'}, \psi_N^{'}, m_N^{'}\right)$, for whom the multipoles $B_n^r\left(\phi_p^{'}, \psi_p^{'}, m_p^{'}\right)$ delete the harmonics $B_n^l\left(\phi_p, \psi_p, m_p\right)$;
    \item we check the total normalized harmonics $b_n\left(\phi_1^{'}, \psi_1^{'}, m_1^{'}, \dots, \phi_N^{'}, \psi_N^{'}, m_N^{'}\right)$ and we stop if $b_2$ and $b_3$ are within few units (the higher order harmonics are already within one unit after the first iteration);
    \item otherwise we update the contributions of the left coil $B_n^l\left(\phi_p^{'}, \psi_p^{'}, m_p^{'}\right)$ and we repeat the steps 2 and 3.
\end{enumerate}

\section{Results}

\begin{figure}[t]
\includegraphics[width=0.4\textwidth]{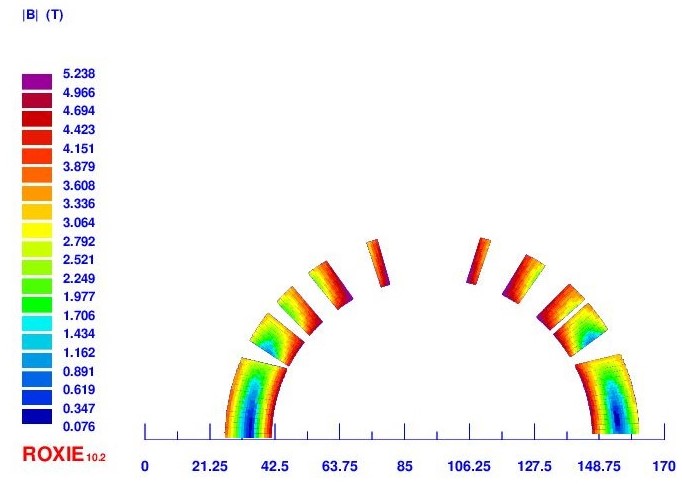}
\caption{Asymmetric D2 coil cross-section.}
\label{fig:D2 current coil}
\end{figure}

\begin{figure}[t]
\includegraphics[width=0.4\textwidth]{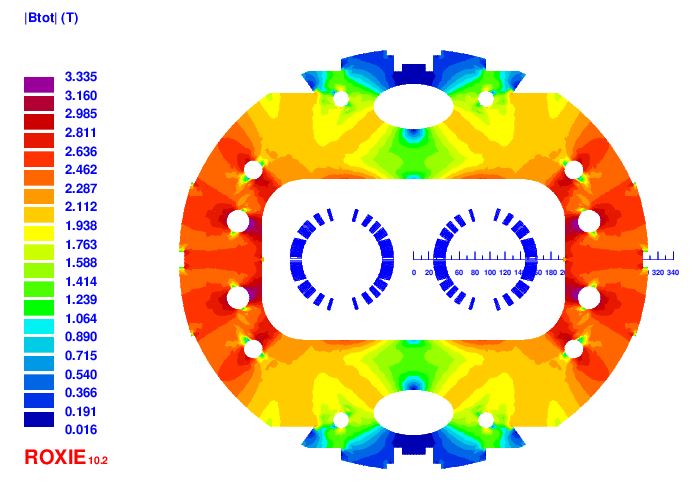}
\caption{Iron yoke of the dipole D2.}
\label{fig:D2 iron yoke}
\end{figure}

The High Luminosity upgrade of LHC requires the replacement of the superconducting magnets, before and after the interaction regions (IR) of the ATLAS and CMS experiments~\cite{Bottura:magnets}. An important role is played by the dipoles recombining and separating the two proton beams around the IR~\cite{Todesco:IR}. These dipoles, D1 and D2, bend the beams in opposite directions. In particular, D2 is a twin-aperture magnet, with an aperture diameter of $105$~\si{\milli\meter} and a inter-beam distance of $188$~\si{\milli\meter}. The dipole must generate an integrated magnetic field of $35$~\si{\tesla\meter} with the same polarity in both apertures. The coils are wound with the same Rutherford cable already used in the outer layer of the LHC bending dipole. The main features of the dipole D2 are listed in Table~\ref{tab:D2 main features}. Fig.~\ref{fig:D2 current coil} shows the asymmetric coil cross-section and Fig.~\ref{fig:D2 iron yoke} shows the optimized shape of the iron yoke.
\begin{table}[ht]
    \caption{Main features of the dipole D2.}
    \begin{tabular}{l c r}
        \toprule
        \toprule
        Feature & Unit & Value \\
        \midrule
        Bore magnetic field & \si{\tesla} & $4.5$ \\
        Magnetic length & \si{\meter} & $7.78$ \\
        Peak field & \si{\tesla} & $5.26$ \\
        Operating current & \si{\kilo\ampere} & $12.34$ \\
        Stored energy & \si{\mega\joule} & $2.28$ \\
        Overall current density & \si{\ampere/mm^2} & $443$ \\
        Magnet physical length & \si{\meter} & $8.11$ \\
        Aperture diameter & \si{\milli\meter} & $105$ \\
        Beam distance & \si{\milli\meter} & $188$ \\
        Operating temperature & \si{\kelvin} & $1.9$ \\
        Operating point on load-line & \% & $66.7$ \\
        Multipole variation due to iron saturation & unit & $<10$ \\
        Number of apertures & & $2$ \\
        \midrule
        Material & & \ce{NbTi} \\
        Cu/Non-Cu & & $1.95$ \\
        No. of strands & & $36$ \\
        Strand diameter & \si{\milli\meter} & $0.825$ \\
        Cable bare width & \si{\milli\meter} & $15.1$ \\
        Cable bare inner thickness & \si{\milli\meter} & $1.362$ \\
        Cable bare outer thickness & \si{\milli\meter} & $1.598$ \\
        Insulation azimuthal thickness & \si{\milli\meter} & $0.1$ \\
        Insulation radial thickness & \si{\milli\meter} & $0.125$ \\
        \bottomrule
        \bottomrule
    \end{tabular}
    \label{tab:D2 main features}
\end{table}

This dipole was designed at INFN in the last years and a short model is currently under construction by ASG Superconductors in Genoa~\cite{Bersani:D2}. The magnetic design was performed by the usual numerical codes, which required a computational time of the order of the seconds to evaluate both the asymmetric coil cross-section, made of $4$ or $5$ blocks of Rutherford cables, and the iron yoke saturation. Because the numerical algorithms took thousands of evaluations, the optimization has been time-consuming.

We reconsidered this design on the basis of the developed analytic approach and searched new coil configurations with four and five asymmetric blocks.

By using the software Wolfram Mathematica 11.3~\cite{Wolfram:Mathematica}, we wrote a code, where the second step of the iterative method is resolved by a differential evolution algorithm.

By performing just one iteration, we minimized the quadratic sum of the $b_n^{coil}$ in Eq.~\eqref{eq:normalized coil harmonics} up to $b_9^{coil}$. In this way we generated a coil configuration to estimate, by a usual numerical code and assuming the iron yoke in Fig.~\ref{fig:D2 iron yoke}, the shifts $\Delta b_n^{sat+geom}$ for each harmonic. We saw that only $\Delta b_2^{sat+geom}$ and $\Delta b_3^{sat+geom}$ were not negligible (about $-200$ units and $-80$ units respectively). Then, by the iterative method we minimized the quadratic sum of Eq.~\eqref{eq:total normalized harmonics} up to $b_9$ for $4$ sectors and $b_{11}$ for $5$ sectors, with the only non-zero terms $\Delta b_2^{sat+geom}$ and $\Delta b_3^{sat+geom}$.

Our code evaluated the asymmetric coil cross-section, made of $4$ or $5$ sectors, in a computational time of about $2$ milliseconds, i.e. about $1000$ times faster than the traditional numerical codes. The computational time of every iteration has been at maximum of about $7$-$8$ minutes. The code stopped after about $2$ iterations. This means that about every $15$ minutes we had the coordinates of a possible asymmetric coil cross-section, which already considered with good approximation the iron yoke saturation (see Tab.~\ref{tab:D2 harmonics}). Thanks to this speed we were able to scan a much higher number of possible configurations and this allowed to find more than $40$ possible solutions with the coil cross-section made of $4$ or $5$ sectors.

\begin{figure}[t]
\includegraphics[width=0.4\textwidth]{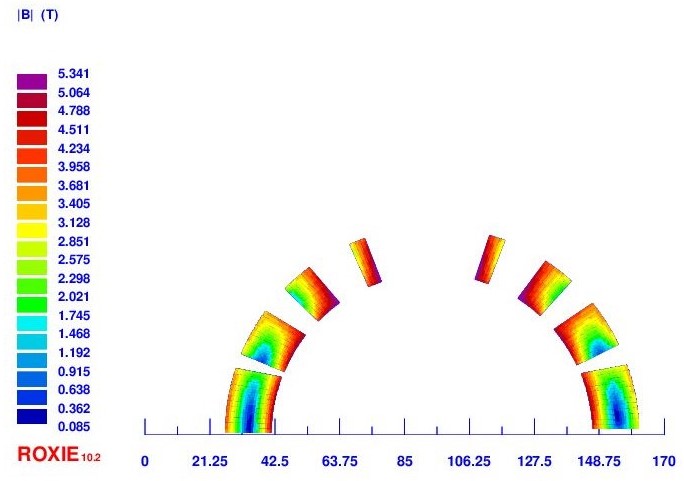}
\caption{Alternative asymmetric D2 coil cross-section.}
\label{fig:D2 alternative coil}
\end{figure}

We inserted in ROXIE~\cite{Russenschuck:ROXIE} the configurations found by our code, for computing the peak fields and the operating margins, with the coil cross-section made of blocks of Rutherford cables and the iron yoke showed in Fig.~\ref{fig:D2 iron yoke}. Fig.~\ref{fig:D2 alternative coil} shows the solution which best fit the specifications and Table~\ref{tab:D2 harmonics} shows the harmonics of this configuration. The first line displays the harmonics at the nominal current, when this solution has been inserted in ROXIE. The second line shows the harmonics after a small fine tuning on the positions and on the tilts of the blocks by means of ROXIE. The current intensity in each block is $12.72$~\si{\kilo\ampere}, the peak field is $5.34$~\si{\tesla} and the operating point on the load-line is about $68.3$\%. This design has a light better field quality than the current one, but it has a light lesser margin on the load-line. These results show that in principle this configuration could be a valid alternative to the current one.
\begin{table*}[t]
    \centering
    \caption{Normal harmonics at operating current for the dipole D2. The first line shows the harmonics of the $4$-block configuration when it has been inserted in ROXIE. The second line displays the harmonics of this configuration after a small fine tuning by means of ROXIE. The last line shows the harmonics of the current $5$-block configuration for a comparison.}
    \begin{tabular}{l c c c c c c c c c c c c c c c c c r}
    \toprule
    \toprule
    $b_{2}$ & $b_{3}$ & $b_{4}$ & $b_{5}$ & $b_{6}$ & $b_{7}$ & $b_{8}$ & $b_{9}$ & $b_{10}$ & $b_{11}$ & $b_{12}$ & $b_{13}$ & $b_{14}$ & $b_{15}$ & $b_{16}$ & $b_{17}$ & $b_{18}$ & $b_{19}$ & $b_{20}$ \\
    \midrule
    $8.98$ & $3.23$ & $9.35$ & $2.92$ & $0.72$ & $-1.81$ & $0.08$ & $-0.66$ & $-0.12$ & $-0.09$ & $-0.2$ & $0.14$ & $-0.29$ & $0.41$ & $-0.71$ & $-1.12$ & $0.12$ & $-0.09$ & $0.09$ \\
    \midrule
    $0$ & $0$ & $0$ & $0$ & $0$ & $0$ & $0.05$ & $0$ & $-0.25$ & $0.03$ & $-0.37$ & $-0.02$ & $-0.38$ & $0.43$ & $-0.7$ & $-1.09$ & $0.11$ & $-0.10$ & $0.09$ \\
    \midrule
    $0$ & $0$ & $0$ & $0$ & $0$ & $0$ & $0$ & $0$ & $0$ & $-1.89$ & $-1.80$ & $-1.91$ & $-1$ & $-0.76$ & $0.12$ & $-0.04$ & $0.14$ & $0.16$ & $-0.07$ \\
    \bottomrule
    \bottomrule
    \end{tabular}
    \label{tab:D2 harmonics}
\end{table*}

\begin{figure}[ht]
\includegraphics[width=0.4\textwidth]{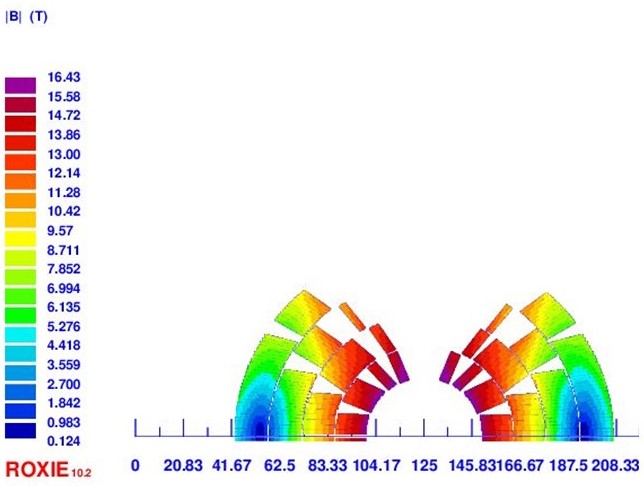}
\caption{Asymmetric coil cross-section for the FCC bending dipole.}
\label{fig:16T current coil}
\end{figure}

\begin{figure}[ht]
\includegraphics[width=0.4\textwidth]{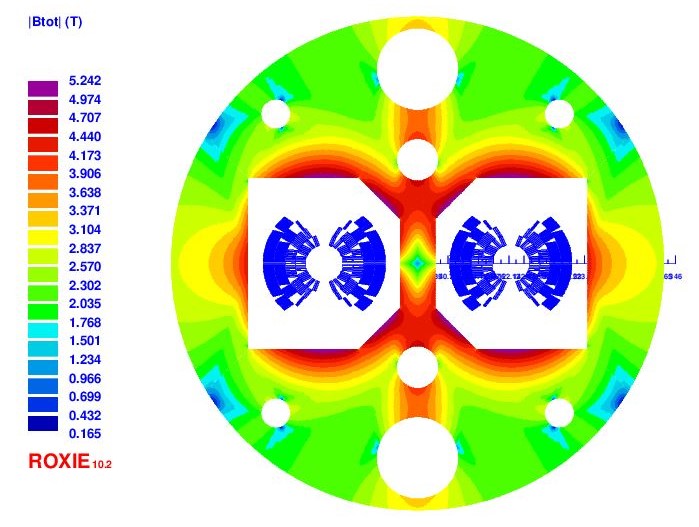}
\caption{Iron yoke of the FCC bending dipole.}
\label{fig:16T iron yoke}
\end{figure}

We applied the analytic method also to the design of $16$-\si{\tesla} bending dipole for the Future Circular Collider. This magnet is a twin-aperture dipole, with an aperture diameter of $50$~\si{\milli\meter} and a inter-beam distance of $250$~\si{\milli\meter}. The main requirements of the dipole are listed in Table~\ref{tab:16T main features}. Fig.~\ref{fig:16T current coil} shows the asymmetric coil cross-section and Fig.~\ref{fig:16T iron yoke} shows the optimized shape of the iron yoke. Each coil is made of two double pancakes, which are connected in series. Each double pancake is wound using its own conductor and this allows the third and fourth layer to have a thinner conductor. This technique is called “grading” and increases the efficiency of the outer layers to produce the main field. The main parameters of the two conductors are reported in Tab.~\ref{tab:16T conductors}. High Field (HF) conductor refers to the first and the second layer, while Low Field (LF) conductor refers to the third and the fourth layer.

The magnetic design was performed by the usual numerical codes, which required a computational time of the order of a few seconds to evaluate both the asymmetric coil cross-section, made of $12$ blocks of Rutherford cables, and the iron yoke saturation. Because the numerical algorithms took tens of thousands of evaluations, the optimization has been time-consuming one again.
\begin{table}[ht]
    \caption{Main design requirements for the FCC dipole.}
    \begin{tabular}{l c r}
        \toprule
        \toprule
        Feature & Unit & Value \\
        \midrule
        Material & & \ce{Nb3Sn} \\
        Bore magnetic field & \si{\tesla} & $16$ \\
        Magnetic length & \si{\meter} & $14.3$ \\
        Aperture diameter & \si{\milli\meter} & $50$ \\
        Beam distance & \si{\milli\meter} & $250$ \\
        Iron yoke outer radius & \si{\milli\meter} & $330$ \\
        Operating temperature & \si{\kelvin} & $1.9$ \\
        Operating point on load-line & \% & $86$ \\
        Cu/non-Cu & & $\geq 0.8$ \\
        Maximum no. of strands & & $40$ \\
        Field harmonics (geom/sat) & unit & $\leq 3/10$ \\
        Number of apertures & & $2$ \\
        \bottomrule
        \bottomrule
    \end{tabular}
    \label{tab:16T main features}
\end{table}
\begin{table}[ht]
    \caption{Main features of the FCC conductors}
    \begin{tabular}{l c c r}
        \toprule
        \toprule
        Feature & Unit & HF & LF \\
        \midrule
        Material & & \ce{Nb3Sn} & \ce{Nb3Sn} \\
        Cu/Non-Cu & & $0.82$ & $2.08$ \\
        No. of strands & & $22$ & $38$ \\
        Strand diameter & \si{\milli\meter} & $1.1$ & $0.7$ \\
        Bare width & \si{\milli\meter} & $13.2$ & $14$ \\
        Bare inner thickness & \si{\milli\meter} & $1.892$ & $1.204$ \\
        Bare outer thickness & \si{\milli\meter} & $2.0072$ & $1.3261$ \\
        Insulation thickness & \si{\milli\meter} & $0.15$ & $0.15$ \\
        Keystone angle & \si{\degree} & $0.5$ & $0.5$ \\
        Operating current & \si{\kilo\ampere} & $11.44$ & $11.44$ \\
        Peak field & \si{\tesla} & $16.4$ & $12.7$ \\
        Operating point on load-line & \% & $86$ & $86$ \\
        \bottomrule
        \bottomrule
    \end{tabular}
    \label{tab:16T conductors}
\end{table}

We applied our analytic approach following the same procedure of D2 and we searched new configurations with a lower or equal number of sectors. By performing just one iteration, we minimized the quadratic sum of the $b_n^{coil}$ in Eq.~\eqref{eq:normalized coil harmonics} up to $b_{10}^{coil}$. In this way we generated a coil configuration to estimate, by a usual numerical code and assuming the iron yoke in Fig.~\ref{fig:16T iron yoke}, the shifts $\Delta b_n^{sat+geom}$ for each harmonic. We considered non-negligible only the terms $\Delta b_2^{sat+geom}$ and $\Delta b_3^{sat+geom}$ (about $30$ units and $-20$ units respectively). Then, by the iterative method we minimized the quadratic sum of Eq.~\eqref{eq:total normalized harmonics} up to $b_{10}$ for $11$-$12$ sectors, with the only non-zero terms $\Delta b_2^{sat+geom}$ and $\Delta b_3^{sat+geom}$. Our code evaluated the asymmetric coil cross-section up to $12$ sectors in a computational time of about $6$ milliseconds, i.e. once again about $1000$ times faster than the traditional numerical codes. The computational time of every iteration has been at maximum of about $20$ minutes. The code stopped almost always after one iteration. This means that about every $20$ minutes we had the coordinates of a possible asymmetric coil cross-section, which already considered with excellent approximation the iron yoke saturation (see Tab.~\ref{tab:16T harmonics}). Thanks to this speed we were able to scan a much higher number of possible configurations and this allowed to find more than $30$ possible solutions with the coil cross-section made of $10$-$12$ sectors.
\begin{table*}[t]
    \centering
    \caption{Normal harmonics at operating current for the $16$-\si{\tesla} bending dipole. The first line shows the harmonics of the new configuration when it has been inserted in ROXIE. The second line displays the harmonics of this configuration after a small fine tuning by means of ROXIE. The last line shows the harmonics of the current configuration for a comparison.}
    \begin{tabular}{l c c c c c c c c c c c c c c c c c r}
    \toprule
    \toprule
    $b_{2}$ & $b_{3}$ & $b_{4}$ & $b_{5}$ & $b_{6}$ & $b_{7}$ & $b_{8}$ & $b_{9}$ & $b_{10}$ & $b_{11}$ & $b_{12}$ & $b_{13}$ & $b_{14}$ & $b_{15}$ & $b_{16}$ & $b_{17}$ & $b_{18}$ & $b_{19}$ & $b_{20}$ \\
    \midrule
    $0.82$ & $-1.68$ & $0.15$ & $-0.66$ & $-0.02$ & $0.17$ & $0.02$ & $-0.36$ & $0.02$ & $1.09$ & $0$ & $-0.26$ & $0$ & $-0.05$ & $0$ & $-0.05$ & $0$ & $0$ & $0$ \\
    \midrule
    $0$ & $0$ & $0$ & $0$ & $0$ & $0.49$ & $0$ & $-0.24$ & $0.02$ & $1.13$ & $0$ & $-0.25$ & $0$ & $-0.05$ & $0$ & $-0.05$ & $0$ & $0$ & $0$ \\
    \midrule
    $0.01$ & $0.22$ & $0.31$ & $0.17$ & $0.35$ & $0.19$ & $0.37$ & $0.57$ & $0.13$ & $1.1$ & $0.09$ & $-0.24$ & $0.03$ & $-0.02$ & $0$ & $-0.06$ & $0$ & $0$ & $0$ \\
    \bottomrule
    \bottomrule
    \end{tabular}
    \label{tab:16T harmonics}
\end{table*}

\begin{figure}[ht]
\includegraphics[width=0.4\textwidth]{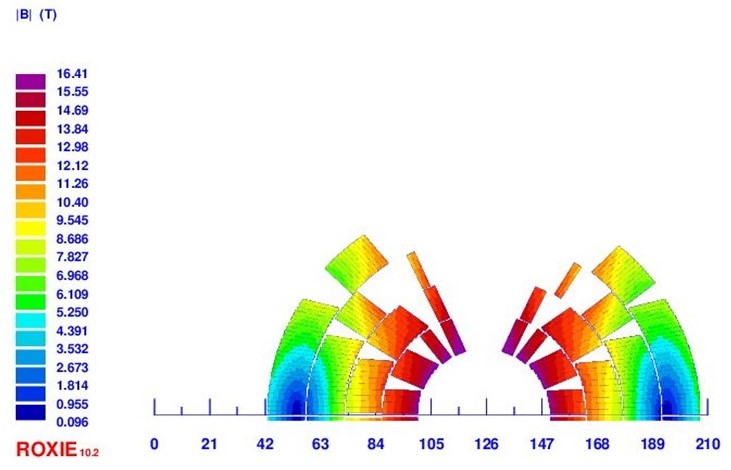}
\caption{Alternative asymmetric coil cross-section for the FCC bending dipole.}
\label{fig:16T alternative coil}
\end{figure}

We inserted in ROXIE~\cite{Russenschuck:ROXIE} the configurations found by our code, for computing the peak fields and the operating margins, with the coil cross-section made of blocks of Rutherford cables and the iron yoke showed in Fig.~\ref{fig:16T iron yoke}. Fig.~\ref{fig:16T alternative coil} shows the solution which best fit the specifications and Table~\ref{tab:16T harmonics} shows the harmonics of this configuration. The first line displays the harmonics at the nominal current, when this solution has been inserted in ROXIE. The second line shows the harmonics after a small fine tuning on the positions and on the tilts of the blocks by means of ROXIE. The current intensity in each block is $11.41$~\si{\kilo\ampere}, the peak fields are $16.4$~\si{\tesla} in the HF conductor and $12.5$~\si{\tesla} in the LF conductor, the operating points on the load-line are about $86$\% for the HF conductor and about $85$\% for the LF conductor. This design has a light better field quality than the current one and it has a light higher margin on the load-line in the LF conductor. These results show that in principle this configuration could be a valid alternative to the current one.

\section{Conclusions}

We developed an extension of the sector model for the magnetic optimization of the twin-aperture $\cos\theta$ superconducting dipoles. It enables to minimize the magnetic cross-talk by finding the asymmetric coil configurations and by considering the iron yoke saturation. This analytic method allows a very fast computation of the field harmonics with respect to the conventional optimization tools (about $10^3$ times faster). The improved speed allows to perform a much lager scan over the possible coil cross-sections and so to increase the possibilities to find the coil layout which best fits the requirements. This method was applied to two different magnets and for both it allowed to find new configurations, which could be a valid alternative to the current ones.

\end{document}